# Fractal Nature and Scaling Exponents
# of Non-Drude Currents in Non-Fermi Liquids

J. C. Phillips

Bell Laboratories, Lucent Technologies (Retired)

Murray Hill, N. J. 07974-0636

USA

## ABSTRACT

In many oxides of the perovskite and pseudoperovskite families there are phase transitions between insulating and normal metallic (Fermi liquid) phases that are separated by an intermediate phase that is often called a non-Fermi liquid (NFL). The dc resistivity of the intermediate or NFL phase often exhibits a $T$ temperature dependence, in contrast to the $T^2$ dependence expected from a "bad" normal metal. The same alloys exhibit a non-Drude (ND) $\omega^{2\alpha}$ frequency dependence, with $\alpha \sim 0.5$, in contrast to the Drude dependence $\omega^2$ characteristic of samples with the $T^2$ behavior. Various attempts have been made to modify the algebra of *continuum* Fermi liquid theory (FLT) to derive the ND exponent $\alpha$, but these have been based on artifices designed to explain only this one parameter. The *discrete* filamentary model has been used to calculate many properties of high temperature superconductors, and to explain the asymmetric nature of the intermediate phase. Here it is used to derive $\alpha$ by the *same rules* previously used for several other discrete relaxation calculations that are in excellent agreement with other quite different experiments. The results are: (cubic) perovskites, $\alpha = 0.45$, and planar conductivity of bilayered pseudoperovskites, $\alpha = 0.70$. The corresponding experimental values are (0.4, 0.5) and 0.7.

## I. INTRODUCTION

The discovery of high-temperature superconductivity (HTSC) (Bednorz and Mueller, 1986) has led further to the discovery of many allied normal-state transport anomalies, which are characteristically found in oxides with the (nearly) cubic perovskite or related (nearly) tetragonal layered pseudoperovskite structures. Some attempts to understand the origin of HTSC, and the closely related phenomena of Colossal Magnetoresistance (CMR), have therefore focused on these latter normal-state transport anomalies. However, so far many theoretical models stress the importance of electron-electron interactions, as summarized in (Metzer *et al.,* 1998; Disertori and Rivasseau, 2000; Farid 2000). These theories have been formulated only in the context of continuum models in momentum space that idealize the materials as "black boxes" (the Effective Medium Approximation, or EMA), with results that closely resemble Fermi liquid theory (FLT), which itself gives a good account of normal metallic transport *without* anomalies. Continuum attempts to discuss the anomalies are thus forced to introduce various artifices, especially with regard to relaxation time anisotropies in momentum space (Ioffe and Millis, 1998), that have no general significance, and amount to contriving as many constructs as there are observables.

The distinction between continuous and discrete structures is fundamental in all theories of transport in solids. In simple metals transport theory is always based on continuum models in momentum space, with the interaction between current carriers and impurities being treated by perturbation theory. However, in the limit $T \rightarrow 0$ near the metal-insulator transition (MIT) in semiconductor impurity bands, I have identified (Phillips, 1999a, 2000a) an intermediate metallic phase in which electronic transport occurs only along discrete, optimized dopant-enriched (locally metallic) filamentary paths in position space. Many scientists have found this network model counter-intuitive, as no prescription can be given for constructing these optimized paths (for much the same reason that the traveling salesman problem is insoluble). However, the distinct topological nature of discrete models leads to many predictions, both qualitative and quantitative, that contradict the expectations of continuum models. These distinct predictions often involve *no adjustable parameters* because they are topological, and so depend on dimensionality alone. The rules for carrying out the calculations are supported both by many experiments, and by extensive numerical simulations on relaxation in classical model glasses. One of the great, and perhaps unexpected, benefits of the theory is that it shows that most of the perturbative paraphenalia of momentum space FLT are irrelevant to the observed anomalies. Considering how large the anomalies are, and how elaborate the paraphenalia, this is a most satisfactory aspect of the theory.

The current experimental situation is summarized in (Dodge *et al.,* 2000), which also contains many experimental references. The simplest and most economical, as well as analytically most appealing, fit to the experimental data is obtained with the relation (van den Marel 1999)

$$\sigma_{ND}* = \sigma_0 \, (\sigma_D \, (\omega,\tau)/\sigma_0)^{\alpha} \qquad (1)$$

$$\sigma_D \, (\omega,\tau) = \sigma_0 \Gamma (1/\tau - i\omega)^{-1} \qquad (2)$$

Thus $\sigma_D \, (\omega,\tau)$ is the Drude conductivity, and the non-Drude behavior of the measured $\sigma_{ND}*$ is contained in the parameter $\alpha$ in Eqn. (1). Experimentally in transition metal oxides that show the T, rather than $T^2$, anomaly in $\rho(T)$, $0 < \alpha < 1$.

In conventional transport theory, which utilizes plane-wave basis states (Boltzmann equation, etc.), all the complexities of the transport processes are contained in the various approximations that are used to calculate the momentum dependence of the relaxation time $\tau$. So it also is with FLT. In materials where the temperature and frequency dependence of $\sigma$ is dominated by impurity, rather than phonon, scattering, both the $T^2$ and the $\omega^2$ dependence predicted by FLT at low T and large $\omega$ are in good agreement with experiment. The theoretical problem now is to find a model that can be used to explain the linear T dependence and can also calculate the parameter $\alpha$ without introducing contrived assumptions with regard to the parameters $\Gamma$ and/or the relaxation time $\tau$.

It turns out that the key to obtaining $\alpha$ lies in a deep understanding of the nature of relaxation, not in traditional Fermi liquid momentum (**q**) space, but rather in an abstract configuration space which contains position (**r**) space as a subspace. In HTSC the nature of this abstract hyperspace is determined by impurities and defects (Phillips, 1987). It is the special configurations assumed in suitably prepared samples by impurities and defects that distinguish weakly metallic and carefully processed transition metal oxides from the ordinary metals for which FLT was originally designed. Just as the transport properties of transition metal oxides differ drastically from those of ordinary metals, so also there are drastic differences between momentum space (essentially relevant to jellium, or perturbative variations on jellium), and this new perovskite-specific configuration space. The characteristics of the new hyperspace depend very sensitively on the nature of the nanoscale order that exists in the dopant configurations. This order shares many essential aspects with the order that is characteristic of good atomic or molecular glasses. In the past it has been customary to describe such nanoscale order by phrases such as (micro-) phase separation, but this phrase is rather vague, and will not be used here.

What do we know about relaxation in glassy systems? Only a few decades ago, both the nanoscale order of glasses and the nature of their relaxation were considered to be quite mysterious. It was customary, for example, to describe network glasses (such as window glass) as "continuous random networks". Today there exists a very successful topological theory of the global structure of network glasses (Thorpe *et al.*, 1999; Selvanathan *et al.*, 1999; Kerner and Phillips, 2001), and an equally successful, and closely related, topological theory of the relaxation of a very wide range of electronic and molecular glasses (Phillips, 1996). It turns out that the topological methods that have been so successful in explaining relaxation in glasses can be applied, almost without change, to derive non-Drude behavior in non-Fermi liquids.

To see how this could happen, it is convenient to summarize what we now know about relaxation in glasses. Suppose the glass is perturbed by an impulse at t = 0, and then is allowed to relax to its initial state. An observable shift $\Delta I(t)$ will then relax in essentially two stages. The first stage is a simple exponential, while the second stage is a stretched exponential. Although relaxation in the second stage is often described as "residual", in typical homogeneous glasses the extrinsic transient initial exponential accounts for only ¼ of the relaxation, which is essentially macroscopic (boundary-condition dependent). Thus the intrinsic microscopic part contributes ~ ¾ of the relaxation. It is described by

$$\Delta X(t) = \Delta X(0) \exp[-(t/\tau(T))^{\beta}] \tag{3}$$

with X ~ 3I/4.  Here $0 < \beta < 1$, and with $\beta = 1$, we recover simple exponential relaxation, which is appropriate to gases and low-viscosity liquids.  Similarly, in (1), if we put $\alpha = 1$, we recover Drude currents, which are appropriate to simple metals, which resemble gases of current carriers.  Because we are dealing with relaxation phenomena in both cases, and because the ways in which the exponents $\alpha$ and $\beta$ enter (1) and (3) are so similar, the reader should not be greatly surprised to learn that the ideas that have been used successfully to calculate $\beta$ can also be used with equal success to calculate $\alpha$.

## II.     WHY AND HOW FLT FAILS

FLT is designed to describe the properties of normal metals with very large residual resistivities.  The $T^2$ term is sometimes explained in terms of the phase space available for scattering of electron pairs.  At the same time momentum conservation tells us that when two electrons with the same effective mass scatter from each other, their current is unchanged.  These effects of conservation laws in FLT have been discussed (Metzer *et al.,* 1998; Disertori and Rivasseau, 2000; Farid 2000) from a perturbative viewpoint, the expansion being made in the strength of the electron-electron Coulomb interaction.  Supposing that the perturbation series were summed successfully, one would still need to examine the quasiparticle-impurity scattering, because this is the source of momentum dissipation.  We can expand these damping processes in powers n of the impurity density (linked clusters with n impurities).  Such an expansion is sketched in Fig. 1.

In Fig. 1(a) we have $(\mathbf{q},\mathbf{q}')$ pair scattering in the absence of an impurity.  In this (jellium) case, because of momentum conservation, there is no damping, unless we add a phenomenological relaxation time to the model, yet the phase space factor $\Omega^2 = |\, (E(\mathbf{q}) - E_F)(E(\mathbf{q}') - E_F\,) \,|$ is present.  At this stage one can add a phenomenological relaxation time to provide damping, but this artifice does not show how we are to deal with more complex cases.  In Fig. 1(b) we have scattering in $\mathbf{r}$ space, and the two electrons are scattering from a cluster containing one impurity.  In order for this interaction to be strong (large residual resistivity), it should be resonant, that is, there should be a large enhancement of the scattering cross section for $\Omega^2 << W^2$, where W is a single-particle band width. Such an enhancement can come from an exchange interaction (electron-hole scattering), with the quasiparticles being resonantly localized near the impurity. Of course, diagrams such as Fig. 1(b) are the basis of FLT which contains chemical information, that is, the dependence of the residual resistivity on the nature of the impurity.

Now suppose we add more and more impurities to the cluster. There are two possibilities: the impurities cluster in a compact way, and the size of the cluster grows like $n^s$, where $s = 1/d$ and d is the dimensionality. This is the case for normal metals, and it is clear that in the presence of compact clusters FLT is still valid. However, there is a second possibility, that the impurities cluster in a singular way, such that the overlap between successive pairs of resonant states on successive impurities still is large, but the overall overlap within a cluster is minimized. This singular, extended configuration can be described as filamentary (Fig. 1(c)). The largest dimension of this singular configuration scales with $s > 1/d$. For a straight filament, $s = 1$. It is clear that in the limit $n \rightarrow \infty$ such filamentary configurations can generate a new phase. Is this phase a possible candidate for the non-FL (NFL) phase? Yes. Is it a candidate with the required physical properties? Yes, as has already been shown for $\rho(T)$ and for the phase diagram (Phillips, 1999b, 2000a), and as will be shown here for $\sigma_{ND}^*(\omega,\tau)$. Is it the only such candidate? Quite probably, but that is difficult to prove.

### III.    HYPERSPACE DESCRIPTION OF ANOMALOUS RELAXATION

Relaxation is a very old subject, whose experimental roots go back to Faraday. The fundamental breakthrough in understanding the microscopic nature of relaxation was made by Arrhenius in his Ph. D. thesis (1884), for which he received the Nobel prize (1903). He observed that the non-equilibrium nature of relaxation implies that it cannot be described merely in terms of energy conservation, but must involve thermal activation over what we now call saddle-point bottlenecks in configuration space, with the activation barrier energy $\Delta E_a$. In molecular glasses, these bottlenecks are well localized in position space (Dzugutov, 1999), which means that  even in Fermion systems momentum space may provide a poor starting point for understanding relaxation processes. Typically $\Delta E_a$ is very large (~ 1 eV), that is, it does not scale with any of the obvious small quasi-particle energies, such as phonons or magnons, but is more comparable to band widths W or bond-breaking energies. At low temperatures, where T $<< \Delta E_a$, the relaxation time may be hours or days, but the fundamental energy scale for the relaxation is always large.

Anomalous relaxation, that is, stretched exponential relaxation, Eqn. (3) with $\beta < 1$, is an historical oddity: it was discovered empirically by Kohlrausch nearly 150 years ago. The microscopic explanation that identified $\beta$ and derived the stretched exponential was not obtained until 120 years later (Scher and Lax, 1973). Even then, however, the story was incomplete, for the Scher-Lax relationship, $\beta = d/(d + 2)$, agrees with only part of the experimental data. To obtain complete agreement, one must construct a hyperspace with dimensionality d*, so that $\beta = d^*/(d^* + 2)$. In general, d* = fd, with $0 < f < 1$, and only in some cases does f = 1. The important point about the fractal or hyperspace construction is that, in spite of the apparent complexity of anomalous relaxation, f can be calculated quite easily (Phillips 1996, 2000b) in cases where the internal structure is simple.

To justify this unexpected remark, the reader might wish to spend a week or two studying the 75-page review article (Phillips 1996), but for those readers who are pressed for time and would prefer a shorter explanation, the example of axial quasicrystals (Elser and Henley, 1985) is most convenient. The hyperspace approach is the basis of Penrose's prediction of the possibility of the existence of crystals with 5-fold symmetry. In the Penrose construction the Euclidean coordinates **r** become $\mathbf{r}_{\parallel}$, and the Penrose projective coordinates are $\mathbf{r}_{\perp}$. The Penrose coordinates $\mathbf{r}_{\perp}$ provide the mathematical constructs which make possible tilings that give rise to five-fold symmetries that are incompatible with normal crystals. Motion in position or $\mathbf{r}_{\parallel}$ space (the first d channels) involves *phonons* and produces relaxation. On the other hand, motion in $\mathbf{r}_{\perp}$ space (the complementary second channels) involves *phasons,* which only *rearrange* particles without diffusion or relaxation. In the ideally random quasi-crystal a given hop occurs in the full Penrose configuration product space, (real$\parallel$) $\otimes$ (complementary $\perp$). Thus it may take place along either $\mathbf{r}_\parallel$ or $\mathbf{r}_\perp$ . This means that the effective dimensionality d* in which diffusive hopping takes place is given by

$$d* = f\,d \qquad\qquad (4)$$

where the fraction f(p) measures the *dimensional effectiveness for actual diffusion* of hopping in pd channels, only d of which are associated with relaxation (Phillips, 1994). For simple isotropic glasses (for example, polymers such as butadiene, which lack bulky side groups) p has nearly always turned out to be either 1 or 2. Although larger values of p, corresponding to larger hyperspaces, are possible in principle, so far they have not been unambiguously observed.

For an axial quasicrystal, which is quasi-periodic only in the plane normal to the axis, the calculation is somewhat more complex. There are five channels, three phonons in $\mathbf{r}_\parallel$ space, but only two phasons in $\mathbf{r}_\perp$ space, so that $f_p = 3/5$, and $d*_p = 9/5$, a rather unusual dimensionality! Thus $\beta_p = d*_p/(\,d*_p + 2) = 9/19$, an even more unusual fraction! However unexpected this peculiar fraction may be, it is in *spectacularly excellent agreement* (Dzugutov and Phillips, 1995) with large-scale MDS, which gave $\beta_p = 0.47$.

If there are any skeptics left at this point who doubt the validity of the fractal projection procedure, it is suggested that they ask themselves two simple questions: (1) Just how often in the theoretical physics of complex systems has anyone ever achieved agreement like this (no adjustable parameters)? (2) Can such agreement really be coincidental? Are the *dozens of other successes based on extremely high-quality data* that originally led to this idea (Phillips, 1994) (*before* the quasicrystal simulations were done) also coincidental? Or is this just a new idea, which requires some getting used to?

## IV. DO FILAMENTS REALLY EXIST?

Before we proceed to the simple derivation of $\alpha$, it is important to discuss just why oxide perovskites and pseudoperovskites are able to exhibit such striking filamentary effects. Isn't it extremely unlikely that the dopants should self-organize into filaments? In general, it is, but in general most materials also do not exhibit HTSC or CMR. This raises the question of whether or not filamentary formation can be *specific* to oxide perovskites and pseudoperovskites. If it is, then the very fact that HTSC and CMR are rare properties is a strong point in favor of the (otherwise unlikely) filamentary model.

Oxides are ionic crystals with strong internal electric fields. Most HTSC are cuprates in which the active dopant is an oxygen vacancy. (LSCO is a special case that will be discussed later.) The structural chemistry of cupric compounds is exceptionally complex; "in one oxidation state this element shows a greater diversity in its stereochemical behavior than any other element" (Wells, 1984). This means that the activation energy for oxygen diffusion can be very small, as there are many sites with nearly equivalent energies. Moreover, the perovskite structure is characteristically ferroelastic, which means that the structure will contain not only strong internal stress fields, but also ferroelastic nanodomains (Phillips, 2000a: Phillips and Jung, 2001). These also contribute to the observed oxygen mobility, which in perovskites is among the largest observed in any crystal class.

We must now consider two contributions to the free energy: (1) the reduction in configurational entropy associated with filament formation, which increases the free energy, and makes the existence of filaments appear to be unlikely, and (2) the higher conductivity of filaments, which provides better screening of fluctuating internal electric fields, thereby reducing the free energy, which favors the formation of filaments. It seems that in many ionic crystals there could be an intermediate temperature range where (2) can outweigh (1), but in most cases the dopant mobility is so low that filamentary configurations are not attainable on laboratory time scales. The high oxygen mobility specific to perovskites thus represents a very important factor that makes filamentary formation possible. Other factors, such as layering into alternately metallic and semiconductive regions, may also be important, but high dopant mobility is probably the most necessary factor.

The foregoing discussion has explored the simplest possible justification for the existence of filaments in the context of materials science, that is, the special properties of oxide perosvikites. This, and neither whimsical conjectures concerning forward scattering in black boxes, nor *ad hoc* toy models, is the appropriate way to approach the theory of HTSC, as the distinguishing feature of the cuprates is that they lie at the cutting edge of materials science. (One may also note that the fundamental theoretical problems posed by HTSC cannot be treated by any kind of perturbation or conventional scattering theory, but instead lie at the cutting edge of the intersection between modern mathematics and theoretical physics, as discussed elsewhere (Phillips, 2001)). There is also a great deal of microscopic evidence that supports the filamentary model, especially in the 50 meV vibronic anomalies found in neutron vibrational spectra (Reichardt *et al.*, 1989) and infrared (especially c-axis) spectra (Homes *et al.*, 1995), also discussed elsewhere (Phillips and Jung, 2001).

## V. FRACTAL DESCRIPTION OF NON-DRUDE CONDUCTIVITY

How can we characterize these filaments topologically? Because of their multinary chemistry and perovskite structures, HTSC and CMR compounds contain a high density, and wide variety, of spatial inhomogeneities (Mueller *et al.*, 1987). One might picture the self-organized filaments as avoiding randomly distributed phase-separated obstacles, but this picture is incomplete because the results could depend on the sizes and densities of the obstacles. In addition to minority phases, these obstacles (or insulating "holes") could be underdoped dead ends or insulating nanodomains The correct approach is to recognize that the intermediate phase lies between the insulating phase, at lower dopant densities, and the Fermi liquid phase, at higher dopant densities. The transition between the insulating phase and the intermediate phase is continuous (Phillips 1999b), and corresponds to what is usually called the metal-insulator transition in continuum models.

One of the characteristic features of perovskite and pseudoperovskite structures is that nearest neighbor bond angles are nearly $\pi/2$ (most cations) or $\pi$ (oxygens), and so these structures can be described well by central force models. For such models it is believed that the percolative threshold and exponent are the same for shear waves and the electrical conductivity (Plischke and Joos, 1998). Thus the percolative model can explain why in LSCO the transition between the insulating state and the intermediate state is marked by the onset of a divergence between the macroscopic orthorhombicity and the local orthorhombicity, as measured in EXAFS experiments (Haskel *et al.*, 1996). This agreement reflects the distinctive ferroelastic nature of Stormer filaments (Phillips, 1999c).

The topological properties of percolating filaments have been studied numerically in great detail. For our present purposes the relevant quantity is the fractal dimension $d_{min}$ of self-organized *minimum paths* (Hermann and Stanley, 1988). (This is very different from the randomly percolating backbone dimension (Janssen *et al*., 1999), which is appropriate to percolation in a network of random resistors.. The use of the latter would produce much poorer agreement between theory and experiment. Indeed, the good agreement with experiment obtained here for *minimum paths* is strong evidence for the self-organized nature of filamentary paths.) The length $l$ of the shortest coherent path or dynamical distance between two points of a strongly disordered material is greater than the Pythagorean distance r. For self-similar or fractal materials, the relationship between $l$ and r is

$$\log l = \text{const.} + d_{min} \log r \qquad (5)$$

where $1 < d_{min}$, and $d_{min} = 2$ for a random walk, which occurs for $d \geq d_c = 6$ in the classical case, and for $d \geq d_q = 4$ in the quantum case. For our purposes, the relevant values of $d_{min}(d)$ are those calculated in the classical case with great precision (Hermann and Stanley, 1988), $d_{min}(2) = 1.130(2)$, and $d_{min}(3) = 1.34 (1)$. For small values of d, one can interpolate between $d_{min}(1) = 1.000$ and $d_{min}(2 )$ to obtain other values of d, for instance $d_{min}(1.5) = 1.06$.

We can now calculate the fractal filling factor f associated with percolating filaments,

$$f(d) = d_{min}(d)/d \qquad (6)$$

If we consider the Drude current in the high-frequency limit, $\omega\tau > 1$, the physical picture is that in general, the applied field is neither parallel nor perpendicular to the local filamentary tangent, which determines the direction of current flow. If we resolve the applied field into its components that are parallel and perpendicular to the local filamentary tangent, then we see that the probability that the field is entirely parallel to the tangent is just f. In fact, f measures the fractal effectiveness of an ac field in contributing to the conductivity, as the perpendicular component simply causes the current to oscillate back and forth in the filament, with no net contribution to the conductivity. (If the minimal path were strictly linear, then f would be just 1/d. Again, use of this linear approximation would significantly reduce agreement between theory and experiment.) This means that one can integrate over the angle between the field and the tangent to obtain (van den Marel 1999)

$$\sigma^* = (\sigma_\perp)^{1-\alpha} (\sigma_{||})^\alpha \qquad (7)$$

Now if we put $\sigma_\perp = \sigma_0$, and $\sigma_{||} = \sigma_D (\omega,\tau)$, we obtain Eqns. (1) and (2). One can elaborate on this derivation by observing that the contribution to $\sigma$ from each carrier is proportional to its mean free path $l$, and that the differential contribution $d\sigma$ of each carrier on each filamentary element is therefore proportional to $dl$. Thus summing over $d\sigma$ is the same as summing over filamentary length elements $dl$, which leads again to (7).

Previous discussions have been carried out in momentum space $\mathbf{q}$ with angular anisotropies dictated by the Fermi surface and the crystal structure (van den Marel 1999). So far it would seem that all that we have done is to show that all such calculations can be carried out for nanoscale-disordered samples in a rotationally invariant manner in $\mathbf{r}$ space, with no reference to either the crystal structure or to the Fermi surface. Certainly this seems to be a much more appealing approach, particularly since it shows the close connection of non-Drude conductivities to the metal-insulator transition and to the unified theory of the phase diagram (Phillips, 1999b) and vibronic anomalies (Phillips and Jung,

2000). However, this is not the end of the story. It is evident from the previous discussion that

$$\alpha = \alpha(d) = f(d) \qquad (8)$$

as given by (6). Thus all that matters in calculating $\alpha$ is the effective dimensionality d of the network, because $d_{min}(d)$ is already known.

For a nearly cubic $ABO_3$ perovskite (A = Sr, B = Ru), it is clear that d = 3 and $\alpha = 0.45$, in good agreement with the experimental estimates (Dodge *et al.*, 2000), $\alpha = 0.4$, and (Kostic, *et al.*, 1998), $\alpha = 0.5$. For YBCO the filamentary model says (Fig. 2) that the filaments pass alternately between the chains (d = 1) and the planes (d = 2), so the only sensible choice for the planar conductivity is d = 1.5. This gives $\alpha = 0.70$, again in excellent agreement with the experimental value of 0.7 (Dodge *et al.*, 2000).

It would be interesting to test the theory further for a case where d = 2. The first possibility that comes to mind is a material with only $CuO_2$ planes, such as LSCO. Unfortunately, the midinfrared conductivity is not well fitted by (1), probably because of the very large midinfrared absorption peak near 0.15 eV (Dolgov *et al.*, 1997). There is, however, a promising alternative. The fractal character of the filamentary metric has been used (Phillips 1998a) to calculate the difference in optimal doping concentrations (Chmaissem et al. 1999) in cation-substituted alloys L*B*CO = $(La_{1-x}Ca_x)(Ba_{1.75-x}La_{0.25+x})Cu_3O_{7\pm\delta}$ compared to $YBCO_{7-\delta}$. These alloys are interesting for several reasons: first, varying x does not change the net cation charge, and so the position of the Fermi energy relative to the density of states peak is not changed by charge transfer from the insulating planes to the cuprate planes (the rigid band or effective medium factor). Second, compared to YBCO, where the oxygen sites in the "CuO" plane are less than half-full (p = 0.46 at maximum $T_c$, here $\delta = 0.18$ and p = 0.59, these sites are more than half-full: why is this necessary? Third, unlike YBCO, the oxygen atoms in the $CuO_{1\pm\delta}$ plane do not segregate into alternating rows or b-axis chains with orthorhombic symmetry and a simple linear metric, but instead are randomly distributed among competing sites with (a,b) tetragonal symmetry. This means that metallic peercolating paths in these planes will not follow linear chain segments, but instead will follow curvilinear paths with the fractal, non-linear metric defined by $d_{min}$. Calculations (Phillips 1998a) based on these concepts explain why extra oxygens are needed to optimize $T_c$ in the "CuO" plane of the cation-substituted L*B*CO alloys compared to YBCO. Thus we can say that for these alloys d = 2, and the predicted value of $\alpha = 0.56$.

The algebra of the fractal relations (5) – (8) is so simple that it is easy to explore different microscopic models to see whether the predicted differences in $\alpha$ are experimentally significant. For example, what would happen if one were to assume that the metric of the filamentary paths were determined quantum mechanically instead of classically? In the quantum case we should use $d_q = 4$ instead of $d_c = 6$, as in the classical case, which implies small changes in $d_{min}$. These can be interpolated by noting that $d_{min}(1) = 1$, and that $d_{min}(d_c$ or $d_q) = 2$. With $d = 1 + \varepsilon$, we obtain $d_{min}, q - 1 = 5(d_{min,c} - 1)/3$, so that $d_{min,q}(2) = 1.22$ and $d_{min,q}(3) = 1.57$. This gives $f_q(2) = 0.61$, a shift of only 9% compared to $f_c(2) = 0.56$. In three dimensions the quantum effects are larger, $f_q(3) = 0.52$ compared to $f_c(3) = 0.45$. Finally, $f_q(1.5) = 0.73$. Although the differences between the classical and quantum values are smaller than the quoted experimental uncertainties (Dodge *et al.,* 2000), the classical percolation paths are consistently in better agreement with the mean experimental values for both the cubic and the layered cases. (A better interpolation scheme increases the disagreement for the quantum case.) This is reasonable, as the filaments are formed at annealing temperatures, which are typically ~ 700K, where quantum coherence effects are expected to be weak.

## VI.    TEMPERATURE DEPENDENCE AND LOW-FREQUENCY CUTOFF

Although Eqn. (1) has been used to fit the frequency dependence of experimental conductivity data (Dodge *et al.,* 2000; van den Marel 1999), it does not provide any guidance to the interpretation of the linear temperature dependence of the dc resistivity. A low-frequency cutoff ~ 20 cm$^{-1}$ is observed (Dodge *et al.,* 2000) at low temperatures, below which the conductivity saturates. The percolative model provides a simple physical interpretation for the origin of the cutoff and its temperature dependence, but first some comments are appropriate concerning the well-known linear temperature dependence of the planar resistivity.

Certainly one of the motivations for choosing $\alpha = 0.5$ is that it leads to $\sigma^*(\omega) \sim \omega$ for large $\omega$, which is dimensionally consistent with a resistivity linear in T. However, the most recent experimental values for the frequency dependence of the planar conductivity, particularly in the layered case, correspond to $\alpha = 0.7$, yet for the temperature dependence one does not observe that the planar resistivity is proportional to anything but T. This means that the physical mechanism that determines the dc resistivity is *fundamentally different* than that for the ac conductivity above the cutoff. This difference cannot be explained by a minimalist approach that treats the sample as a "black box", governed only by greatly enhanced forward scattering.

The linear temperature dependence was first observed (Xiao *et al.*, 1988) in a Bi-Sr-Cu-O sample with a low $T_c$, so that the resistivity was linear in T from 7K to 700K. Shortly thereafter the Hall number was also shown (Stormer *et al.,* 1988) to be linear; both linearities were explained by the presence of a narrow, high mobility band (the Stormer band) of extended states centered on $E_F$ which are *entirely disjoint* from bands of localized states at more distant energies. It is the purity of these high-mobility states, reordered in energy relative to Fermi liquid states, that one optimizes when one anneals to maximize $T_c$; this same purity is necessary to explain all the normal-state transport anomalies. At the time this model was criticized as microscopically arbitrary, because the width $W_H$ required for the narrow band is $\sim 0.05$ eV, which is much smaller than electronic band widths $W \sim 1$ eV. Later it was shown (Phillips, 1997, 1998a) that this narrow width is naturally associated with the width of the resonant tunneling centers in semiconductive layers (such as BaO) that connect partially metallic layers (the $CuO_x$ planes and chains in YBCO) to form optimized filaments. Alternatively, it is also comparable to pseudogap energies. These filaments are self-organized to maximize the normal-state conductivity near the annealing temperature, thereby minimizing the free energy by screening the internal ionic electric fields.

Here to discuss $\sigma_{ND}*(\omega,\tau)$ we have assumed that the topologies of the self-organized filaments are optimized to have the smallest fractal dimension $d_{min}(d)$ consistent with d-dimensional percolation. Is this *variational* condition equivalent to maximizing the normal-state conductivity near the annealing temperature? It is, because the shorter the path, the lower the resistivity. Another way of emphasizing the significance of $d_{min}(d)$ is to compare the choice of f = d*/d with d* = $d_{min}(d)$ to d* = 1 (a non-fractal straight line) and d* = $d_{back}$, the dimensionality of the backbone of a random resistor network. This is done in Fig. 3. Although the difference between $d_{min}$ and 1 is small, experiment clearly favors $d_{min}$, while d* = $d_{back}$ is clearly inadequate: to obtain good agreement with experiment, the network, although fractally disordered, must be self-organized.

In order to form the Stormer band, electronic energy levels near $E_F$ must be reordered in a non-perturbative way. Such reordering is facilitated by the tendency of defect states in semiconductors to adjust their energies to pin $E_F$. The reader is also surely familiar with the non-perturbative Coulomb-driven reordering (Efros and Shklovskii, 1975) that occurs in semiconductor impurity bands and that creates the pseudogap that is a suitable starting point for understanding pseudogaps in HTSC. Here the reordering exploits the availability of lattice relaxation modes to lower the energy. However, this relaxation is not in the nature of a Jahn-Teller (gap-producing) distortion, but can instead be described as an *anti-Jahn-Teller effect* (Phillips, 1993) that pins high-mobility states in a narrow band at the Fermi energy.

The low-frequency region where the resistivity is linear in T, because of the existence of the Stormer band, could be consistent with the high-frequency region, where $\sigma_{ND}*(\omega,\tau) \sim \omega^{2\alpha}$ only if $\alpha = \frac{1}{2}$ (which is not usually the case). Instead, a crossover occurs near $\omega_c \sim 30$ cm$^{-1}$ (Dodge *et al.,* 2000). Within the percolative context, it is easy to understand (Phillips, 1998b) the origin of this crossover. The magnitude of $\omega_c \sim 30$ cm$^{-1}$ can be understood if we assume that the construction which leads the filamentary paths through the dopant centers involves not only resonant electronic states, but also local vibronic modes. These local vibronic modes include at least one very soft mode that is associated with high oxygen mobility and electronic energy level reordering to form the Stormer band.  It is not easy to observe such low-frequency modes by neutron scattering, and because the filamentary density is so low, they would tend to disappear in the acoustic mode background.  However, such a low-frequency mode is plausible on chemical grounds, as it is known (Martin, 1970) in compounds such as Cu(I, Br, Cl) that one  of the short-range shear moduli is ~ 10% of the shear modulus found in Si and Ge.  More specifically, a low-lying optical vibrational branch has been observed by neutron scattering in α-AgI at 2 meV (Buhrer and Bruesch 1975).  This mode is essential to the phase transition at 147C that forms a superionic conductor.

## VII.    SELF-ORGANIZATION

The strong correlation between the linear temperature dependence of the resistivity and optimization of HTSC has been widely accepted as proof that the linearity is an intrinsic property that should be explained by microscopic theory.  However, it can be argued (Sundqvist and Anderson, 1990) that such linearity is an artifact, as theory describes the system at constant volume, whereas experiments are done at constant pressure.  An estimate of the constant volume effect shows sublinear deviations of order 25%.  It has been suggested from one fit to the data (Dolgov *et al.*, 1997)  that the sublinear deviation could arise in LSCO from the (probably Sr-related) dopant band near 0.15 eV.  However, in cases like YBCO where the dopants are O, it is not clear that such an explanation is valid.

The model of self-organized percolation suggests a different explanation.  Because the samples are measured at constant pressure, the filaments are self-organized at constant pressure.  In other words, the theory is *not* a constant-volume theory, it is a constant-pressure theory, and the observed  linearity is indeed intrinsic at constant pressure.  This general argument does not require specific assumptions about the role of dopant bands.

## VIII.    CONCLUSIONS

NFL behavior in perovskites and pseudoperovskites is closely associated with HTSC and CMR. Hitherto almost all attempts to explain NFL behavior, including ND conductivity, have utilized continuum models in momentum space that are quite similar to FLT itself. Here a discrete approach in position space, which is the logical alternative to the continuum one in momentum space, has been developed. It gives excellent agreement with experiment, without introducing artificial assumptions, and uses no adjustable parameters. The discrete filamentary character is essential to understanding many other features of these materials, including the nature of the intermediate phase and its properties.

FIGURE CAPTIONS

Fig. 1.  (a)  Electron-electron scattering (here between an  electron and a hole, with the Coulomb interactions indicated by dashed lines) cannot alter currents, because of momentum conservation. (b)  When impurities are added to provide a momentum sink, and to explain the residual resistivity, one obtains  results consistent with experiment for many "dirty" metals, including a $T^2$ contribution to the resistivity, and conventional Drude conductivity frequency $\omega^{-2}$ dependence at high frequencies.  In conventional FLT the impurity distribution is disordered (nearly random).  (c) Here the impurities or defects are self-organized, and an intermediate phase with non-Drude conductivity characterized by the frequency $\omega^{-2\alpha}$ dependence at high frequencies.

Fig. 2.  In HTSC the filamentary paths pass alternately between "metallic" planes through resonant tunneling centers in intervening semiconductive planes (Phillips, 1990).  The sharp bends in the filamentary paths (Phillips and Jung, 2001) explain many anomalies in vibronic neutron and infrared spectra.  Because the paths alternate between chains and planes, the best value of d(YBCO) = 1.5.

Fig. 3.  A plot of f = d*/d for three choices of d*:  a non-fractal straight line; the chemically shortest path $d_{min}$ of self-organized minimum paths (Hermann and Stanley, 1988), and the randomly percolating backbone dimension $d_{back}$ (Janssen *et al.*, 1999).  The arrow A is the value for YBCO, while the arrows B and C refer to the $SrRuO_3$ values, as measured in (Kostic *et al.,*1998) and (Dodge *et al.,*2000), respectively.  It is clear that the choices d* = $d_{min}$, d($SrRuO_3$) = 3, and d(YBCO) = 1.5 give the best agreement with experiment.

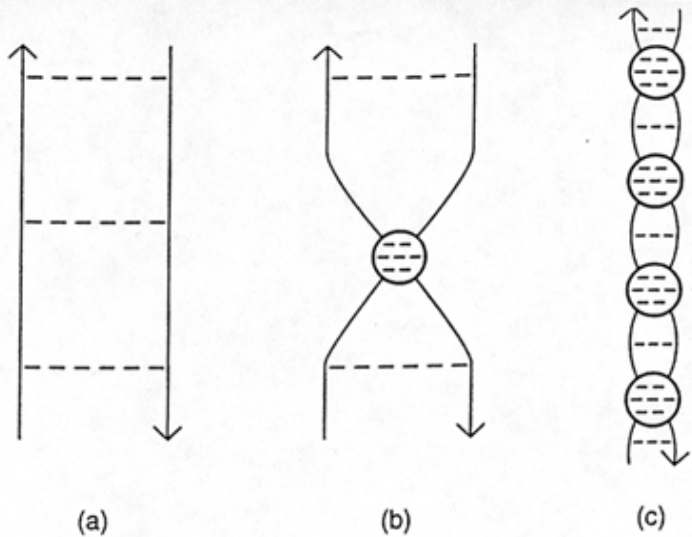

(a)   (b)   (c)

Fig. 1

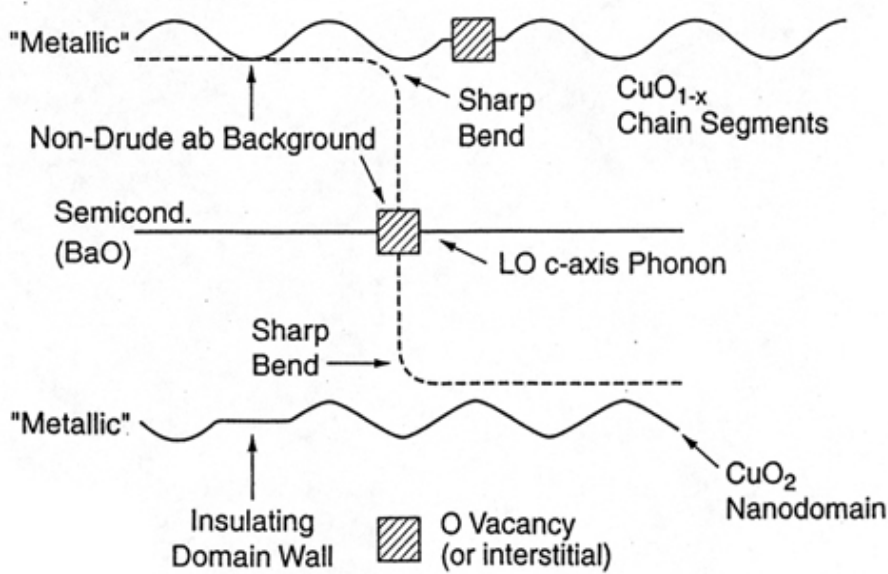

"Metallic" — Sharp Bend — $CuO_{1-x}$ Chain Segments

Non-Drude ab Background

Semicond. (BaO) — LO c-axis Phonon

Sharp Bend

"Metallic" — $CuO_2$ Nanodomain

Insulating Domain Wall     O Vacancy (or interstitial)

Fig. 2

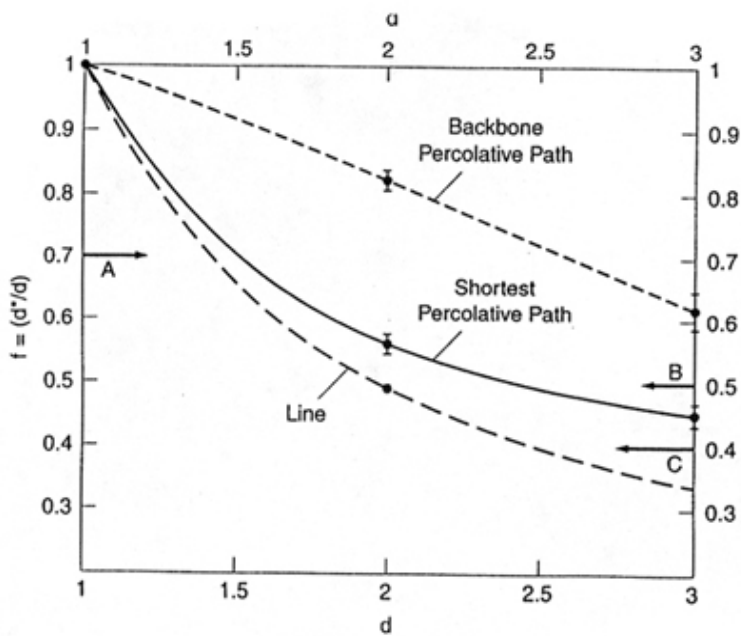

Fig. 3